\documentstyle[prl,aps,twocolumn]{revtex}
\begin{document}
\title{Uniqueness in Relativistic Hydrodynamics} 
\author{Peter Kost\"adt and Mario Liu~\cite{email}}
\address{Institut f\"ur Theoretische Physik, Universit\"at Hannover,\\
         30167 Hannover, Germany\\
         (\today)\\[0.3cm]
\begin{minipage}{14.1cm}\rm\quad
The choice of the 4-velocity in the relativistic hydrodynamics proposed by 
Landau and Lifshitz is demonstrated to be the only one satisfying all 
general principles. Especially the far more widely adopted Eckart choice has 
to be abandoned.\\[0.4cm]
\tt ITP-UH-11/96 \vspace*{-0.3cm}
\end{minipage}}
\maketitle
Three reasons come to mind for seeking a covariant formulation of the 
hydrodynamic theory for simple fluids: In a fundamental vein, we want 
hydrodynamics as a basic theory to be covariant; in fact, this was one of 
the first few tasks tackled after the birth of special relativity. 
More practically, relativistic hydrodynamics is increasingly employed in 
cosmology and astrophysics to study dissipative processes, such as the 
relaxation of inhomogeneities in the early universe, or its viscosity-driven, 
inflationary expansion~\cite{BreHe}. Finally, any residual ambiguities in 
the relativistic hydrodynamics are an indication of our less-than-perfect 
grasp also of the Galilean version. They require close scrutiny. 

Covariant hydrodynamics can be found in most textbooks on relativity and
astrophysics. Better ones give two versions, one due to Eckart~\cite{Eck,Wein},
the other by Landau and Lifshitz (LL)~\cite{LL6}. While Eckart pins the 
macroscopic 4-velocity $u^\mu$ to the total particle current, LL set it 
proportional to the total energy flux. So there is no dissipative particle 
current in the Eckart version, in direct analogy to the non-relativistic case;
while LL --- at first sight somewhat odd --- do have one, 
and they refrain from a dissipative energy current instead. In the literature,
the Eckart version is much more widely employed~\cite{BreHe,Wein,Dixon}, 
it seems the more traditional theory. Yet, both reduce to the familiar 
non-relativistic hydrodynamics for $c\to\infty$; besides, standard textbooks 
(eg. Weinberg~\cite{Wein}) regard the two versions as equivalent, as being 
related by a simple transformation of the velocity. Clearly, one need not 
worry about a mere difference in the reference frame. 

This equivalence, however, is a fallacy. Let us recall how the velocity is 
defined in hydrodynamic theories, consider first the non-relativistic case. 
The standard Gibbs relation, valid not only when equilibrium reigns, takes 
the entropy density $s$ as a function of five conserved densities, energy 
$\epsilon$, momentum $\mbox{\boldmath$g$}$, and mass $\varrho$, 
\begin{equation}
T\,ds=d\epsilon-{\bf v}\!\cdot\!d\mbox{\boldmath$g$}-\mu\,d\varrho\;.
\label{CGFG}\end{equation}
It states unequivocally that the velocity is a thermodynamic quantity, 
${\bf v}\equiv -T(\partial s/\partial\mbox{\boldmath$g$})$, known if the 
local entropy density is. It contains only equilibrium information. 
This is of course the concept of local equilibrium, one of the few founding 
principles of the hydrodynamic theory: It takes far less time to establish 
equilibrium locally than globally; the first is a microscopic time $\tau$  
(referred to as the collision time in dilute systems), the other grows with 
the system dimension and is macroscopic. As long as the frequency is small 
enough, $\omega\tau\ll1$, the Gibbs relation holds, and all thermodynamic 
variables $\epsilon$, $\mbox{\boldmath$g$}$, $\varrho$, and their conjugate 
variables $T$, ${\bf v}$, $\mu$ contain only information about the local 
equilibrium state. Especially, they possess a well defined parity under time 
reversal. The relativistic description is hardly different: The velocity 
${\bf v}$ becomes the 4-velocity, $u_\mu\equiv-T(\partial s^\nu/\partial 
T^{\mu\nu})$, with $s^\mu$ and $T^{\mu\nu}$ being the equilibrium entropy 
4-vector and energy-momentum tensor, respectively. So $u^\mu$ too contains 
only equilibrium information. 

Now, the difference between the two 4-velocities mentioned above is clearly 
in dissipative quantities, so at least one contains non-equilibrium 
information and cannot be the correct hydrodynamic velocity. This argument 
agrees with the prima facie evidence that the two versions of relativistic 
hydrodynamics have different types of differential equations. For example, 
the equation for the transverse velocity is elliptic in the Eckart version, 
and parabolic in that of LL. (In the so-called extended thermodynamic 
theories --- cf. discussion below --- the equation can be rendered hyperbolic 
for either choice of the 4-velocity. Still, as these are extensions of  
different hydrodynamic theories, they remain distinctly different.)

Having clarified that both versions are inequivalent, we obviously need to 
address the question: Which, if any, is the correct theory for relativistic 
dissipative fluids? In seeking uniqueness in a relativistic theory --- since 
the $c\to\infty$ limit proves inconclusive --- it is natural to examine how 
uniqueness is achieved in the Galilean version. Unfortunately, this is 
something of a red herring, as the lack of a dissipative particle current in 
the Galilean hydrodynamics, or ${\bf j}=\varrho\,{\bf v}$, is more a 
statement of microscopic plausibility, and maybe the bold summary of 
countless experiments; it is not the result of a cogent and general 
deduction. In fact, a classic paper by Dzyaloshinskii and Volovik~\cite{Vol} 
proposes to include the  dissipative term 
${\bf j}-\varrho\,{\bf v}\sim\mbox{\boldmath$\nabla$}\mu$. 

Nevertheless, a footnote by LL, in \S49 of \cite{LL6}, states
that the mass current ${\bf j}$ should always be equal to the momentum 
density $\mbox{\boldmath$g$}=\varrho\,{\bf v}$, and therefore cannot possess 
any dissipative terms. Their line of argument, regrettably, falls short of 
being ironclad: Starting from the continuity equation and the center-of-mass 
motion, they assert the validity of $\int dV\,\mbox{\boldmath$g$}=\int 
dV\,{\bf j}$, where the integration volume is that of the total system. The 
reader is left wondering especially about the alleged equivalence to the 
local relation $\mbox{\boldmath$g$}={\bf j}$. 

In this paper, we shall ameliorate both aspects. We provide a clear cut 
proof of $\mbox{\boldmath$g$}={\bf j}$, demonstrating the rigorous validity 
of the standard form of the Galilean hydrodynamics; and we demonstrate that 
only the LL version of the relativistic hydrodynamic theory conforms to all 
general principles. The proof $\mbox{\boldmath$g$}={\bf j}$ takes place in 
very much the same way as deducing the symmetry of the momentum flux, or 
stress tensor, from local angular momentum conservation. The relevant 
conserved density here is ${\bf i}\equiv\varrho\,\mbox{\boldmath$x$}
-\mbox{\boldmath$g$}\,t$, with $\mbox{\boldmath$x$}$ and $t$ denoting 
the space and time coordinates, respectively. 
While ${\bf i}$ is known to be additive, conserved, and the 
direct consequence of the invariance under Galilean boosts~\cite{Doug}, it 
has not been hitherto included in thermodynamic and hydrodynamic 
considerations. As will become clear soon, this is a serious omission. Its 
inclusion not only establishes the form of the mass flux, but also leads to 
general thermodynamic equilibrium conditions that are valid for any 
reference frame. Surprisingly, these simple yet fairly fundamental relations 
are new.  

Relativistically, the information of $\mbox{\boldmath$g$}={\bf j}$ is 
automatically included in the symmetry of the energy-momentum tensor, 
though ${\bf j}$ is now the inertial mass current, ie. the total energy flux 
including the rest mass. Since the momentum density remains a thermodynamic 
variable, with a negative parity under time inversion, neither the momentum 
density nor the inertial mass current may contain any dissipative terms (in 
the local rest-frame). This excludes any covariant theory that does not 
adopt the LL 4-velocity. 

Two pieces of information were needed in each of the above cases: (i) The 
equality of the momentum density to the (rest or inertial) mass current; 
(ii) the fact that the momentum density is a thermodynamic variable with a 
well defined parity. It is ironic that while the condensed matter people 
were wondering about the first piece, which for the relativists is a trivial 
consequence of the 4-notation, the latter ignored the second, something the 
former group never does. We physicists are indeed a community divided by 
notations. Piercing both pieces together, the kinship to be expected between 
the two versions of hydrodynamics becomes evident. For instance, repeating 
the relativistic mistake in the non-relativistic theory, ie. violating the 
second condition while upholding the first, leads to, as will become clear 
soon, 
\begin{equation}\label{NonRelQuatsch}
 \mbox{\boldmath$g$}={\bf j}=\varrho\,{\bf v}
                      -\chi\,(\mbox{\boldmath$\nabla$}\mu+\partial_t{\bf v}). 
\end{equation}
This is in striking similarity to the momentum density in the Eckart 
theory, the dissipative part of which has the form 
$\tau^{i4}=-\chi\,(T^{-1}\partial_i T+\partial_t u_i)$\cite{Wein}. Yet 
Eq.~(\ref{NonRelQuatsch}) is manifestly unphysical: The total and conserved 
momentum must remain $\int d^3\!x\,\varrho{\bf v}$, irrespective whether the 
system is in equilibrium or not, or what its acceleration is. 

Of the three issues plaguing the relativistic hydrodynamics --- uniqueness, 
causality and stability --- we focus on the first. But we need to comment on 
the other two, as they have been the starting points of worthwhile efforts 
in the past that partially tie in with our results. First, causality. 
Strictly speaking, the diffusion equation implies signals with infinite 
velocity, or horizontal ``world lines''. While unphysical generally, this 
defect is aggravated in relativity: When viewed from a different frame, 
the world lines tilt, implying signals that go backwards in time. To repair 
this, extended thermodynamic theories\cite{Dixon,IsrSt} were put forward 
which start from the hydrodynamic theories but include additional dynamic 
variables. The resultant larger set of coefficients can be chosen such that 
all the differential equations are hyperbolic, ensuring causality. The price 
for this nice feature is a rather more complicated theory, and the difficulty 
of finding a universally valid and accepted set of additional variables --- 
except perhaps in dilute systems. 

But we may also take a more perspective view, and accept that the diffusion 
equation is not an exact mathematical statement. Rather, it is an  
approximative description --- confined to the hydrodynamic regime, with an 
accuracy of the order of thermal fluctuations.  Taking this into account, 
(eg. considering only amplitudes of the variables that are above a minimal 
threshold,) the signal velocity never exceeds that of the constituent 
particles~\cite{GLW}, excluding any acausal consequences. 

Next, stability, first in the fluid's rest frame: The LL theory is stable 
with respect to small fluctuations around an equilibrium configuration, not 
so the Eckart version\cite{HisLi}, and remarkably, nor the non-relativistic 
theory that contains Eq.~(\ref{NonRelQuatsch}). In fact, both suffer from 
the same problem. Consider a small but spatially homogeneous velocity field 
with $\mbox{\boldmath$\nabla$}p,\mbox{\boldmath$\nabla$}\mu=0$, the 
Navier-Stokes equation reduces to $\partial_t\mbox{\boldmath$g$}=0$, 
or $\varrho\,\partial_t{\bf v}-\chi\,\partial_t^2{\bf v}=0$, which (in 
addition to the usual ${\bf v}=const$) obviously also contains the run-away 
solution $\sim e^{(\varrho/\chi)t}$. Similarly, with a momentum density that 
contains the acceleration $\partial_t u_i$, the Eckart choice cannot help to 
avoid an analogous instable solution. 

This would represent an independent argument favoring the LL choice, except 
that --- as observed by Hiscock and Lindblom~\cite{HisLi} ---  in frames 
moving with respect to the fluid the diffusion equations in the LL theory 
also develop diverging solutions, which grow exponentially with 
microscopically short characteristic times. For lack of space, we briefly 
summerize our reasons for believing that this frame-dependent instability 
does not constitute sufficient ground to reject the LL choice, and promise a 
detailed account in a forthcoming paper. Consider the parabolic diffusion 
equation, $\partial_t\vartheta-\alpha\,\partial_x^2\vartheta=0$. Its 
characteristics are the lines $t=const$, and only if the initial values are 
prescribed on one of these, do we have a single bound mode, $\delta\vartheta\,
e^{ikx-\alpha k^2t}$, with $k\in{\relax{\rm I\kern-.18em R}}$~\cite{Mik}. 
Initial data on a non-characteristic curve, say $x+\beta\,t=const$,
$\beta\in{\relax{\rm I\kern-.18em R}}$, generally produce two 
independent solutions. For the simplest case of $\beta=0$, they are 
$\delta\vartheta_1\,e^{-i(\Omega x+\omega t)}e^{\Omega x}$ for $x<0$ and 
$\delta\vartheta_2\,e^{i(\Omega x-\omega t)}e^{-\Omega x}$ for $x>0$, 
with $\Omega\equiv(\omega/2\alpha)^{1/2}$, $\omega\in{\relax{\rm I\kern-.18em 
R}}$. In the respective wrong region, one solution appears unbound.
Being invariant with respect to coordinate transformations, the 
characteristics of the boosted diffusion equation that Hiscock and Lindblom 
consider are $t=\gamma(\tilde t+v\,\tilde x)=const$, with $t$ the proper time. 
The solutions they examine, however, satisfy initial data on the 
non-characteristic $\tilde t=const$, where $\tilde t$ is the 
time in the moving frame. So the appearance of an unbound solution for 
$\tilde t\to\infty$ is a mathematical consequence to be expected. 
Nevertheless, the diverging mode, being absent in the rest frame, must not 
be observable in a moving one. And it is not, as it only exists for negative 
times $\tilde t$, and decays for $\tilde t\to-\infty$ within a microscopically
brief period that is outside the hydrodynamic regime. In fact, this mode is 
just one of those signals discussed above that run backwards in time in moving
frames. (These arguments do not apply to the Eckart instability. It happens 
in the fluid's rest frame, where any deviation from the non-relativistic 
hydrodynamics is worrisome.) 

The extended theories are stable for both choices of the 4-velocity if 
linearized; though the Eckart version turns instable again if non-linear 
terms are included~\cite{HisOl}.

We conclude: Within its range of validity, the relativistic hydrodynamics 
is just as healthy as the non-relativistic theory. If someone is willing to 
put up with a few acausal consequences, blatant but recognizably outside this 
range, he retains the benefit of a simpler theory. If not, he may resort to 
the extended theory --- though it has to be one that adheres to the LL choice 
of the 4-velocity.

Let us now consider the hydrodynamics in greater details, starting again with 
the non-relativistic version. The equations of motion for the thermodynamic 
variables of Eq.~(\ref{CGFG}) are, 
\begin{eqnarray} 
&&\partial_ts+\mbox{\boldmath$\nabla$}\!\cdot\!{\bf f}=R/T\,,\quad 
  \partial_t\epsilon+\mbox{\boldmath$\nabla$}\!\cdot\!{\bf q}=0\,,
  \label{EM1}\\
&&\partial_tg_i+\partial_k\Pi_{ik}=0\,,\quad 
  \partial_t\varrho+\mbox{\boldmath$\nabla$}\!\cdot\!{\bf j}=0\,.\label{EM2}
\end{eqnarray}
We explicitly include the conserved quantity 
\begin{equation} 
{\bf I}\equiv{\textstyle\int}d^3\!x\,(\varrho\,\mbox{\boldmath$x$}
               -\mbox{\boldmath$g$}\,t)=M\,{\bf X}(t)-{\bf G}\,t  
\end{equation}
in our consideration, where $M$, ${\bf G}$, and ${\bf X}(t)$ denote the total 
mass, the total momentum, and the center-of-mass coordinate, respectively. 
Clearly, ${\bf I}/M$ is the initial coordinate of the center of mass, so we 
may perhaps refer to ${\bf I}$ as the center-of-mass inertial coordinate 
({\sc comic}), and to ${\bf i}\equiv\varrho\,\mbox{\boldmath$x$}
-\mbox{\boldmath$g$}\,t$ as the {\sc comic} density. 

Neither the angular momentum nor the {\sc comic} requires an independent 
equation of motion. Writing $\partial_t(\varepsilon_{ikm}\,x_k\,g_m)= 
-\partial_n(\varepsilon_{ikm}\,x_k\,\Pi_{mn})+\varepsilon_{ikm}\,\Pi_{mk}$,
one finds that the angular momentum density $\varepsilon_{ikm}\,x_k\,g_m$ 
obeys a continuity equation only if $\Pi_{ik}=\Pi_{ki}$. Analogously,  
$\partial_t(\varrho\,x_i-g_i\,t)=-\partial_k(j_k\,x_i-\Pi_{ik}\,t)+j_i-g_i$ 
holds for the {\sc comic} density, a locally conserved quantity, hence 
$\mbox{\boldmath$g$}={\bf j}$. This concludes the clear cut and simple 
proof we were looking for. 

Next we deduce thermodynamic equilibrium conditions including the 
conservation of the {\sc comic} ${\bf I}$. Maximizing the total entropy 
$S={\textstyle\int}d^3\!x\,s$ subject to the conservation of energy, 
momentum, mass, angular momentum, and {\sc comic}, we have
${\textstyle\int}d^3\!x\,\{\delta s-\lambda_1\,\delta\epsilon+{\bf\Lambda_1}
\!\cdot\!\delta\mbox{\boldmath$g$}+\lambda_2\,\delta\varrho+{\bf\Lambda_2}
\!\cdot\!\delta(\mbox{\boldmath$x$}\times\mbox{\boldmath$g$})-{\bf\Lambda_3}
\!\cdot\!\delta(\varrho\,\mbox{\boldmath$x$}-\mbox{\boldmath$g$}\,t)\}=0$, 
where the eleven coefficients $\lambda_{1,2}$ and ${\bf\Lambda_{1,2,3}}$ are 
constant Lagrange parameters. Employing Eq.~(\ref{CGFG}), we 
deduce, for arbitrary variations $\delta\epsilon$, $\delta\varrho$ and
$\delta\mbox{\boldmath$g$}$ (with $\delta\mbox{\boldmath$x$}=\delta t=0$),
\begin{mathletters}\label{CEC}\begin{eqnarray} 
&& 1/T=\lambda_1,\quad 
   \mu/T=\lambda_2-{\bf\Lambda_3}\!\cdot\!\mbox{\boldmath$x$},\label{CEC2}\\
&& {\bf v}/T={\bf\Lambda_1}+{\bf\Lambda_2}\times\mbox{\boldmath$x$}
             +{\bf\Lambda_3}\,t.
\label{CEC3}\end{eqnarray}
The last expression does not imply an accelerating momentum, as Dixon 
concluded in Ch.4 \S 4d of Ref.~\cite{Dixon}. To see this directly, 
consider uniform space and time translations: Setting now
$\delta\mbox{\boldmath$x$},\,\delta t=const$, and requiring that the 
equilibrium conditions remain unaltered, we arrive at 
\begin{equation}
M{\bf\Lambda_3}=-{\bf\Lambda_2}\times{\bf G}
\end{equation}\end{mathletters} 
and the dependent ${\bf\Lambda_3\!\cdot\!G}=0$. Together, the equilibrium 
conditions~(\ref{CEC}) are explicitly Galilean covariant: Introducing the 
chemical potential $\mu_0=\mu+{1\over2}v^2$ of the local rest-frame, they 
can be expressed as $T=\bar T$, $\mu_0=\bar\mu+{1\over2}
[{\bf\Omega}\times(\mbox{\boldmath$x$}-{\bf X})]^2$, ${\bf v}={\bf V}+
{\bf\Omega}\times(\mbox{\boldmath$x$}-{\bf X})$,
with ${\bf X}={\bf X}(t)$ being the center-of-mass coordinate, and $\bar T$, 
$\bar\mu$, ${\bf V}$, ${\bf\Omega}$ redefined constants. Clearly, $\mu_0$ 
(and hence the density $\varrho$) only depends on the rotation velocity in 
the center-of-mass frame, and not on the center-of-mass motion. 
Without including the {\sc comic} ${\bf I}$ (ie. setting ${\bf\Lambda_3}=0$ 
above) LL obtained, in a similar calculation~\cite{LL5}, 
${\bf v}={\bf V}+{\bf\Omega}\times\mbox{\boldmath$x$}$, and concluded that 
the equilibrium velocity ${\bf v}$ of a general frame has to 
be a constant of time. But this is clearly only correct in special frames, 
when ${\bf V\,\|\,\Omega}$. 

Now Eq.~(\ref{NonRelQuatsch}) is derived. First we remark that the unusual 
form of the thermodynamic force $\mbox{\boldmath$\nabla$}\mu
+\partial_t{\bf v}$ is a natural consequence of Eqs.~(\ref{CEC}): This 
combination vanishes in equilibrium and may therefore serve as a legitimate 
thermodynamic force. More technically, given the existence of 
${\bf j}^{\scriptscriptstyle D}={\bf j}-\varrho\,{\bf v}$, both 
$\mbox{\boldmath$g$}$ and $\epsilon$ acquire a dissipative part, 
$\mbox{\boldmath$g$}=\mbox{\boldmath$g$}^{\scriptscriptstyle Eq}+{\bf j}
^{\scriptscriptstyle D}$ and $\epsilon=\epsilon^{\scriptscriptstyle Eq}
+{\bf v\cdot j}^{\scriptscriptstyle D}$. Substituting Eqs.~(\ref{EM1}) 
and (\ref{EM2}) into the Gibbs relation, 
$T\,\partial_ts=\partial_t\epsilon^{\scriptscriptstyle Eq}-{\bf v}
\cdot\partial_t\mbox{\boldmath$g$}^{\scriptscriptstyle Eq}
-\mu\,\partial_t\varrho$, one obtains the entropy production $R=-{\bf j}
^{\scriptscriptstyle D}\!\cdot(\mbox{\boldmath$\nabla$}\mu+\partial_t{\bf v})
+\cdots$, from which Eq.~(\ref{NonRelQuatsch}) [with $\chi>0$, such that 
$R>0$] results. 

Turning our attention to special relativity, the relevant hydrodynamic
equations~(\ref{CGFG}), (\ref{EM1}), and (\ref{EM2}) generalize to 
\begin{eqnarray} 
T\,ds^\mu=-u_\nu\,dT^{\nu\mu}-\mu\,dn^\mu\,, \label{GR}\\
\partial_\mu(s^\mu+\sigma^\mu)=R/T\,, \label{BE1}\\  
\partial_\nu(T^{\mu\nu}+\tau^{\mu\nu})=0\,,\ 
\partial_\mu(n^\mu+\nu^\mu)=0\,.\label{BE2}  
\end{eqnarray} 
Notations: Greek indices run from 1 to 4, Latin indices only to 3; the speed
of light is unity; the metric is $\eta^{\mu\nu}={\rm diag}(1,\,1,\,1,\,-1)$; 
the coordinate 4-vector is $x^\mu=(\mbox{\boldmath$x$},\,t)$, so 
$\partial_\mu=(\mbox{\boldmath$\nabla$},\,\partial_{t})$; the 4-velocity is 
$u^\mu=\gamma({\bf v},\,1)$ with $\gamma\equiv(1-v^2)^{-1/2}$, hence 
$u^\mu\,u_\mu=-1$;  $\sigma^{\mu}$, $\tau^{\mu\nu}$, and $\nu^{\mu}$ are 
the respective dissipative parts of the entropy 4-flux, energy-momentum 
tensor, and particle 4-flux, they have a different parity under time 
reversal from their reactive counterparts and vanish in equilibrium. 

In the laboratory frame, in which the local fluid velocity is ${\bf v}$, the 
reactive, equilibrium terms are: $s^\mu=su^\mu$, $T^{\mu\nu}=(e+p)
u^\mu u^\nu+p\,\eta^{\mu\nu}$, and $n^\mu=nu^\mu$, where $p=-e+Ts+\mu\,n$ 
is the pressure, $e=\epsilon_0+\varrho$ the density of total energy, 
$n=\varrho/m$ the particle number density, and $s$ as before the entropy 
density, all taken from a local comoving frame. The last three are related by 
$Tds=de-\mu\,dn$, so $\mu=m(\partial\epsilon_0/\partial\varrho)+m$ has a 
different definition than in the Galilean case, but $p$ does not. The 
relativistic version of the Gibbs relation, Eq.~(\ref{GR}), is obtained by 
multiplying the rest-frame relation with $u^\mu$. 
The conservation of the 4-angular momentum is ensured by the symmetry of 
the energy-momentum tensor: $T^{\mu\nu}=T^{\nu\mu}$ and 
$\tau^{\mu\nu}=\tau^{\nu\mu}$. As discussed, this includes the equality of 
the momentum density and the total energy flux. 

Now consider the explicit form of the dissipative terms $\sigma^\mu$, 
$\tau^{\mu\nu}$, and $\nu^\mu$. They are determined by the rate of entropy 
production $R$, a positive Lorentz scalar. Inserting Eqs.~(\ref{BE1}) and
(\ref{BE2}) in (\ref{GR}), and requiring $R$ to be a sum of products of 
thermodynamic fluxes and forces, we arrive at 
$T\sigma^\mu=-u_\nu\,\tau^{\nu\mu}-\mu\,\nu^\mu$ and  
\begin{equation}
R/T=-\tau^{\mu\nu}\,\partial_{(\mu}[u_{\nu)}/T]-\nu^\mu\partial_\mu(\mu/T)
\,,\label{R1}\end{equation}
where the bracket denotes symmetrization, eg. 
$\partial_{(\mu}u_{\nu)}\equiv(\partial_\mu u_\nu+\partial_\nu u_\mu)/2$. 
Global equilibrium conditions are met if the two forces 
$\partial_{(\mu}[u_{\nu)}/T]$ and $\partial_\mu(\mu/T)$ vanish. Irreversible 
thermodynamics generally prescribes the Onsager ansatz, setting 
$\tau^{\mu\nu}$ and $\nu^\mu$ as linear combinations of 
$\partial_{(\mu}[u_{\nu)}/T]$ and $\partial_\mu(\mu/T)$ --- subject to the 
requirements that thermodynamic variables do not possess any dissipative 
counterparts in the local rest-frame ({\sc lrf}). The thermodynamic 
variables of Eq.~(\ref{GR}) reduce to $s^4=s$, 
$T^{\mu4}=(\mbox{\boldmath$g$},\,e)$, and $n^4=n$ in the {\sc lrf}. So 
the lack of dissipative counterparts implies $\sigma^4=\tau^{\mu4}=\nu^4=0$, 
of which the covariant expressions are,
\begin{equation} 
u_\mu\,\sigma^\mu=u_\nu\,\tau^{\mu\nu}=u_\mu\,\nu^\mu=0\,.\label{leqc}
\end{equation}
These are the conditions implemented by LL, and the ones we need to heed 
while evaluating Eq.~(\ref{R1}). 

It must have been a source of confusion that $\mbox{\boldmath$g$}$ itself 
vanishes in the {\sc lrf} --- reducing $T^{\mu4}$ to $T^{44}=e$, 
and seemingly leaving only $\tau^{44}=0$. This overinterprets the {\sc lrf}. 
What we actually need is to examine the infinitesimal changes of the 
variables, $dT^{\mu4}=(d\mbox{\boldmath$g$},\,de)$, and understand that the 
{\sc lrf} does not imply $d\mbox{\boldmath$g$}=0$, as we must allow for 
$\partial_t\mbox{\boldmath$g$},\,\partial_ig_k\not=0$. Non-relativistically, 
of course, $\mbox{\boldmath$g$}$ being a thermodynamic variable is never 
disputed. 

It is not incidental that the conditions (\ref{leqc}) rule out any time 
derivative in $R$. Violating (\ref{leqc}), we find from Eq.~(\ref{R1}):
$\tau^{i4}\sim\partial_t u_i$, $\tau^{44}\sim\partial_t T$, 
$\nu^4\sim\partial_t(\mu/T)$ in the {\sc lrf}. In the equations of motion, 
each yields its own run-away solution, altogether five. Eckart's conditions, 
$u_\mu\,\sigma^\mu=u_\mu\,u_\nu\,\tau^{\mu\nu}=\nu^\mu=0$, partly violate 
Eqs.~(\ref{leqc}), so it is not surprising that his ``momentum of heat''
$\tau^{i4}\sim\partial_t u_i$ gives rise to the instable solution discussed
above. 

The covariant hydrodynamic theory that entails the 4-velocity as proposed 
by LL is the appropriate theory to employ if the velocity difference in the 
system is no longer small when compared to light velocity. However, it does 
not consider charges and electric currents that are frequently present in 
astrophysical systems. For this, one needs in addition the covariant version 
of the dissipative Maxwell equations~\cite{liu}, to be published elsewhere.

We acknowledge financial support of the Deutsche Forschungsgemeinschaft.

\end{document}